\documentstyle[epsf,12pt]{article}
\renewcommand{\section}[1]{\vspace{6pt} \noindent\mbox{#1} \newline \noindent}
\renewcommand{\subsection}[1]{\vspace{6pt} \noindent\mbox{\underline{#1}} 
\newline \noindent}
\renewcommand{\subsubsection}[1]{\vspace{6pt} \noindent\mbox{\underline{#1}}
\noindent}
\newfont{\sansb}{cmssbx10}
\newfont{\sans}{cmss10}

\def\ra{secondary acceleration~}
\def\sp{secondary to primary~}
\def\cry{cosmic ray~}

\def\crys{cosmic rays~}
\def\snr{supernova remnants~}

\def\sn{supernova~}
\def\tcr{\tau_{cr}}
\def\cm{\hbox{cm}}

\def\g{\hbox{g}}
\def\apj{Ap.J.}
\def\et{et.al.}
\def\kms{km/s}

\begin{document}
{\small OG 8.2 \vspace{-24pt}\\}     
\medskip
{\center SECONDARY ACCELERATION OF COSMIC RAYS BY SUPERNOVA SHOCKS
\vspace{6pt}\\}
A. Wandel$^1$ \vspace{6pt}\\
{\it $^1$ Racah Inst. of Physics, The Hebrew University, Jerusalem 91904, Israel\\}
{\center ABSTRACT\\}
In the common model supernova shock-acceleration of cosmic rays
there are two open questions: 1. where does the high energy cosmic rays below
the knee (10$^4-10^6$ Gev) come from, and 2. are cosmic ray  accelerated only
at their origin or contineuosly during their residence in the Galaxy.
We show that $10^15$ eV light nuclei are probably
 accelerted by associations of supernovae.
The ratio of the spectra of secondary to primary cosmic rays would be 
affected by repeated (or secondary) acceleration in the
ISM during their propagation in the galaxy.
The observed secondary and primary CR spectra are used to constrain the
amount of such \ra by supernova remnants (SNR).
Two cases are considered:
weak shocks ($1<M<2$) of old, dispersed remnants, and
strong shocks ($M>3$) of relatively young remnants.
It is shown that weak shocks produce more \ra than what is permitted
in the framework of the standard leaky box  (SLB) model,
making it inconsistent with dispersed acceleration that should be produced 
by SNR.
If the SLB is modified to allow a moderate amount of RA by week
shocks, the RA produced by old SNRs agrees with the rate reqired
to fit the secondary-to primaray cosmic-ray data, making a self consistent 
picture.
Significant \ra  by strong shocks of young SNRs
should lead to flattening of the secondary-to primaray ratio at
high energies, near 1TeV/nucleon.

\setlength{\parindent}{1cm}

\section{WHAT IS SECONDARY ACCELERATION?}
The \cry spectrum in the range 1-10$^5$ is probably produced by shock acceleration in 
the Galaxy (e.g. Axford 1981). 
Its power-law shape $J(p)\sim p^{2.7}$ is beleived to be produced by strong shocks, 
probably of young SNR (Blandford and Ostriker, 1980), 
combined with escape from the Galaxy. After the \cry particle
has left the SNR site and is propagating in the Galaxy, it may enconter another SNR,
be trapped and be accelerated to a higher energy. However, the probability to 
encounter a SNR is proportional to its volume. It turns out to be significant only 
for old SNR, hens with weak shocks.
Indeed, Observations of the ratio of primary to seconday \crys suggest such a \ra process,
but at the same time constrain its amount (Eichler 1980, Cowsick 1986).
Wande et.al. (1987) have calculated the \cry spectrum and the constraints on \ra
set by the data under various \ra models, with the \ra amount and the shock-strength
as parameters. Applying these models to RA by SNR (Wandel 1987;1988) shows that 
one cannot avoid \ra if one assumes that the primary \crys are produced by supernovae:
the young SNR producing the the \crys will expand and occupy a large enough fraction
of the Galaxy to contribute a considerable amount of \ra . The standard leaky-box
model has therefore to be modified to include RA by SNR.

\section{BEYOND THE KNEE}
Acceleration by supernova shocks can produce \crys with energy up to 10$^{15}$ eV. 
It is difficult to reach higher energies, because when the gyroradius of the 
accelerated particles becomes of the order of the shock size, the shock-acceleration
mechanism cannot function, as the particle escapes. 
The gyroradius is given by
\begin{equation}
\label{Rg}
R=\frac{p}{eZB}\sim (10pc)\frac{E_{15}}{ZB(\mu G)},
\end{equation}
so that the maximal energy is
\begin{equation}
\label{Eg}
E_{15}= 0.1R(pc) ZB(\mu G).
\end{equation}
where $E_{15}$ is the maximal energy in units of 10$^{15}$ eV.
To reach energies beyond 10$^{15}$ eV, larger shocks are required, and those can be 
obtained by going to older SNR and to associations of supernovae. 
Older SNR have weaker shockes, which produce a steaper power law.
Indeed, a significant steepening is observed in the \cry spectrum beyond 10$^{15}$.
This could indicate that old SNR associate to produce shocked volumes of sizes
of 100pc and more. 
With galactic magnetic fields of 3$\mu$G 100pc shocks, if  abundant enough,
could produce the observed \cry spectrum up to 3 10$^{16}$eV for protons and up to
10$^{18}$eV for iron. The formalism presented by Wandel (1988) is used to 
convolve the size of the SNR with teir age and shock strength, to determine the
probability and rate of \ra, and the efffective slope of the produced \cry spectrum.

\section{THE SECONDARY ACCELERATION MODEL}

The steady state distribution of a particle population subject to contineous 
acceleration during its propagation is described by the integral equation 
\begin{equation}
\label{J}
J_o(p)-(R+S)J(p)=B(q)\left [ J(p)- (q-1)\int_{p_o}^p \frac {dx}{x}\left (\frac{x}{p}
\right )^q
J(x)\right ]
\end{equation}
where $J_o$ is the source distribution,$R=R_0(p/p_o)^{-\alpha}$ and S are loss terms 
due to escape from the
galaxy and spallation, B is the rate of encountering accelerating events and
\begin{equation}q=\frac{2M^2+2}{M^2-1}
 \end{equation}
is the index of the power law ($J\sim p^{-q}$) produced by the shock of Mach number M
(q=2 for strong shocks, and larger for weaker shoks).
The distribution of each secondary species satisfies a similar equation, with the 
source term replaced by the sum over all primary species $J_i$ which contribute to the
secondary in question, $J_s(p)=\Sigma S_{is}J_i(p)$, with $S_{is}$ is the spallation 
rate from i to s, and $J_i$ satisfies equation (3).

\section{SECONDARY ACCELERATION AND THE LEAKY BOX MODEL}

In the framework of the Standard Leaky Box (SLB) model
no \ra is allowed, except the margin due to the observational
uncertainties.
If the SLB model is modified by allowing a modest amount of \ra
(and assuming a larger escape rate,
which compensates for the increase in the \sp ratio
caused by \ra ), the constraint on \ra is
less stringent.
Wandel et.al. (1987) have calculated the \sp (boron/carbon, denoted by B/C)
energy spectrum for a modified leaky box
models (MLB) with various escape
laws, \ra rates and shock indices $q$.
Their best fit of the B/C data permits a
significantly larger \ra rate than in the SLB model.
For shocks with $q=4~~(M=1.7)$
\begin{equation}
\cases { B< 0.03 & $R_0=0.11$  \quad (SLB) \cr
B\approx (0.2\pm0.05) &
$R_0=0.2$ \quad (MLB)\cr}
\end{equation}
where $B$ and $R_0$ are measured in units of inverse path length
$(\g\cm^{-2})^{-1}$.

Similar constraints can be derived from the primary spectra of
protons and alpha particles (e.g. Webber, Golden and Stephens 1987).

\section{ACCELERATION BY SUPERNOVA REMNANTS}

\vspace{-18pt} \subsection{Weak shocks}

The number of \snr encountered by a \cry particle during its
residence time in the galactic disk,
($\tcr\sim 10^7$yr at a few GeV/nucleon) 
is given by
$N_{cr}
={4\pi\over 3}r^3 S \tcr (1+C)^3 ,$
where $S$ is the \sn rate (per unit volume) in the galaxy, and $C=r_{cr}(t)/r$ represents the \cry
propagation during the remnant's lifetime. The average \cry residence time
in the galactic disk can be written in the form $\tcr =
(1.3c m_p n R_0)^{-1}\approx 5 10^5 R_0^{-1}\hbox{yr}. $ The reacceleration rate
due to \snr becomes
\begin{equation} B(q)=
N_{cr}R_0=
3.0~S_{12} r_{q2}^3\left [1+ {r_{cr}(t_q)\over r_q}
\right ] ^3
=0.14~S_{12} n^{-1} u_{q2}^{-2}(1+C_q)^3 ,
\end{equation}
where  $u=dr/dt= 100 u_2 \kms$ is the shock velocity,
the shock index $q$ is related to the Mach number $M=u/c_s$
by $q(M)$ and
$S_{12}=S/ 10^{-12}$ pc$^{-3}\hbox{yr}^{-1}$
(one \sn every 30 years corresponds to $S_{12}=0.5$).
We take the sound speed in the ISM  $c_s=150\kms$, so $M=1.7 (q=4)$
corresponds to $u=250 \kms$.
If the effective density seen by \snr is that of the warm component
of the ISM,
$n_{eff}=0.1$ (McKee and Ostriker 1977), then $r_q\sim 23$pc,
$C_q\sim 0.2-0.4$ (depending on the \cry propagation, Wandel 1988),
and eq. (6) gives $B=0.2-0.4$.
While this figure is by a factor 10 larger than the constraint in
the SLB model, it is just the \ra rate required by the modified
model in order to fit the B/C data.

\subsection{Strong-Shock Secondary Acceleration by Young SNRs}

The effect of \ra by young SNRs on the \sp ratio has been considered in
previous papers (Wandel 1990; 1991). Although the probability of encountering
a young SNR is small, its effect on the \cry distribution at high energies 
is strong, and may be detected.
At high energies the solutions for the \cry distribution (eq. 3)
become assymptotically power laws, with the power index depending
on the source acceleration index $q_0 $, the \ra index $q$, and the
escape law $R\propto p^{-\alpha}$.
It can be shown that the \sp ratio has the
assymptotic form (Wandel et.al. 1987)
\begin{equation}
{J_s\over J_p} \rightarrow \cases
{p^{-\alpha} & $q\geq q_0+\alpha$\cr
p^{-(q-q_0)} & $q<q_0+\alpha $\cr } 
\end{equation}
In particular, for $q=q_0$ the \sp ratio should flatten at high
energies. The energy at which this flattening occurs depends on the
\ra rate $B$  and on the \ra index $q$. 
From the observed \cry spectra we have
$q_0+\alpha\sim 2.7$ and $\alpha\sim 0.5-0.6$.
A significant flattening of the B/C ratio requires shocks with $q\le 2.4$,
 which gives $M>4$ and $u=600 \kms$.
Young \snr may evaporate the cold phase of the ISM
(McKee and Ostriker 1977), in which case $n_{eff}=1$.
If this is not the case, only the warm component is swept by the shock,
so $n_{eff}=0.1$.
Finally, If the
\sn expands within a cavity of hot ISM, $n_{eff}=0.003-0.01$.
Eq. (6) gives
$$B(q\leq 2.4)=
\cases {0.003  &$n=1$  \hbox{(evaporation)}\cr
0.03 & $n=0.1$  \hbox{(no~ evaporation)}\cr
0.3 & $n=0.01$  \hbox{(hot~ ISM)}\cr }$$
If the transition between evaporative and non-evaporative
expansion happens at an intermediate velocity (Cowie, McKee and
Ostriker 1981), $B$ will be between 0.003 and 0.03.
The \sp (Boron/Carbon) ratio has been observed
up to $\sim$1 TeV by the Spacelab-2 experiment of the University of
Chicago (Meyer \et 1987).
The poor statistics at the high energy end prevent a decisive conclusion.
However, preliminary data  suggest that
the B/C ratio in the 0.1-1TeV range is approximately 0.1.
Comparing this with the theoretical
calculation we find that such
a B/C ratio requires a \ra rate  $B(q=2-3)\sim 0.03-0.1$, so
strong-shock \ra is probably important, and young \snr see a relatively
low effective density.
Even if the B/C ratio beyond 100GeV/n is eventually found to decrease,
strong shock \ra by evaporative
young \snr places a lower limit on the \ra rate at $B=0.003$, which
would predict a flattening near 1TeV.

\section{REFERENCES}
\setlength{\parindent}{-5mm}
\begin{list}{}{\topsep 0pt \partopsep 0pt \itemsep 0pt \leftmargin 5mm
\parsep 0pt \itemindent -5mm}
\vspace{-15pt}
\item
Axford, W.I. 1981,  {\it Proc. 17th ICRC (Paris)}.
\item
Blandford, R.D. and Ostriker, J.P. 1980, \apj 237 793.
\item
Cowsik, R. 1986, A\& A 155, 344.
\item
Eichler, D.S. 1980, \apj 237 809.
\item
Engelmann, J.J., \et 1983, {\it Proc. 18th ICRC}
(Bangalore), {\bf 2}, 17.
\item
Meyer, P. \et 1987, {\it Proc. 20th ICRC} (Moscow), 1, 338.
\item
McKee, C.F., and Ostriker, J.P. 1977 \apj 218 148.
%
\item
Wandel, A., Eichler, D.S., Letaw, J.R., Silberberg, R., and Tsao, C.H.
1987 , \apj 317 277.
\item
Wandel, A. 1987, Proc. IAU Colloquium 101 {\it
"Interaction of \snr with the ISM"}, eds. T.L. Landecker, R.S. Roger,
Cambridge University Press.
\item
Wandel, A. 1988, A\& A  200, 279.
\item
Wandel, A. 1990{\it Proc. 21th ICRC} OG 8.2.9.
\item
Wandel, A. 1991{\it Proc. 22th ICRC} {\bf 2}, 237.
\item  Webber, W.R., Golden, R.L., and Stephens, S.A. 1987,
{\it Proc. 20th ICRC} {\bf 1}, 325.

\end{list}

\end{document}